\documentclass{PoS}

\usepackage{heppennames2}

\usepackage{lineno}

\newcommand{\ttbar}{\ensuremath{\PQt\PAQt}\xspace} 
\newcommand{\tch}{\ensuremath{\HepProcess{\PQt \to \PQc \PH}}\xspace}
\newcommand{\tcg}{\ensuremath{\HepProcess{\PQt \to \PQc \PGg}}\xspace}
\newcommand{\tcx}{\ensuremath{\HepProcess{\PQt \to \PQc \HepParticle{E}\!\!\!\!\!\!\!\slash}}\xspace}
\newcommand{\hbb}{\ensuremath{\HepProcess{\PH \to \PQb \PAQb}}\xspace}

\newcommand{\BR}{\ensuremath{\text{BR}}\xspace}
\newcommand{\CLs}{\ensuremath{\mathrm{CL}_\mathrm{s}}\xspace}

\newcommand{\geant}{\textsc{Geant4}\xspace}
\newcommand{\pandora}{\textsc{PandoraPFA}\xspace}
\newcommand{\fastjet}{\textsc{FastJet}\xspace}
\newcommand{\lcfiplus}{\textsc{LcfiPlus}\xspace}

\newcommand{\eettg}{\ensuremath{\Pep\Pem\!\to\PQt\PAQt\PGg}\xspace}

\graphicspath{{./plots/}}

\title{Top-quark physics at the first CLIC stage}

\ShortTitle{Top-quark physics at the first CLIC stage}

\author{
  \speaker{Aleksander Filip \.Zarnecki} ~~~  
        on behalf of the CLICdp Collaboration \\
        Faculty of Physics, University of Warsaw\\
        E-mail: \email{Filip.Zarnecki@fuw.edu.pl}
}

\abstract{
  The Compact Linear Collider (CLIC) is a mature option for a future
  electron-positron collider operating at centre-of-mass energies of
  up to 3\,TeV.
  CLIC will be built and operated in a staged approach
  with three centre-of-mass energies  currently assumed to be
  380\,GeV, 1.5\,TeV and 3\,TeV.
  The energy of the initial stage was chosen to optimize the
  physics potential in terms of Higgs-boson and top-quark measurements. 
  This contribution discusses the prospects for precision measurements
  of top-quark properties at the first stage of CLIC,
  based on detailed simulation studies, taking into account
  luminosity spectra and beam induced backgrounds, full detector
  simulation based on \geant, final state reconstruction based on
  particle flow approach with \pandora, jet clustering with the VLC
  algorithm as implemented in the \fastjet package, and flavour tagging with
  \lcfiplus.

  Based on a dedicated centre-of-mass energy scan around the
  top-quark pair production threshold, the top-quark mass can be
  determined with a precision of about 50\,MeV in a theoretically
  well-defined manner.
  This scan is also sensitive to the top-quark width and Yukawa
  coupling. 
  Other approaches to extract the top-quark mass at the first stage of
  CLIC make use of ISR photons or the direct reconstruction of the 
  top quarks.
  Precise measurements of the differential top-quark pair production
  cross sections at 380\,GeV, for different electron beam
  polarisations, allow the study of  top-quark couplings to
  electroweak gauge bosons.
  Expected limits on new physics contributions described in terms of
  Effective Field Theory (EFT) operator coefficients are
  presented, showing sensitivity of the first CLIC stage to mass
  scales beyond 10\,TeV. 
  The large number of top-quark pairs produced also allows  competitive 
  searches for Flavour Changing Neutral Current (FCNC)
  top-quark decays with charm quarks in the final state.
  Exclusion limits expected for 500\,fb$^{-1}$ collected at
  the first stage of CLIC are presented for \tch, \tcg and \tcx
  channels,
  reaching down to $4.7 \cdot 10^{-5}$ for \BR(\tcg).
  
}

\FullConference{39th International Conference on High Energy Physics\\
		4-11 July 2018\\
		Seoul, Korea}

\begin{document}

\section{Introduction}

Processes involving top quarks provide us unique opportunities to test the
Standard Model (SM) predictions and look for possible signatures of
new physics beyond the SM (BSM).   
Thorough studies of the top-quark physics potential of CLIC, based
on  detector-level simulations, have been recently
presented by the CLIC Detector and Physics
collaboration~\cite{Abramowicz:2018rjq}. 
Summarized in this contribution are the main top-quark physics results
expected after the initial stage of CLIC operation at 380\,GeV.
Prospects for top-quark physics at high-energy CLIC stages
are covered in \cite{ichep_top2}. 
%

\section{Top-quark mass measurements}

\begin{figure}[p]
\begin{minipage}{0.48\linewidth}\centering
\includegraphics[width=\textwidth]{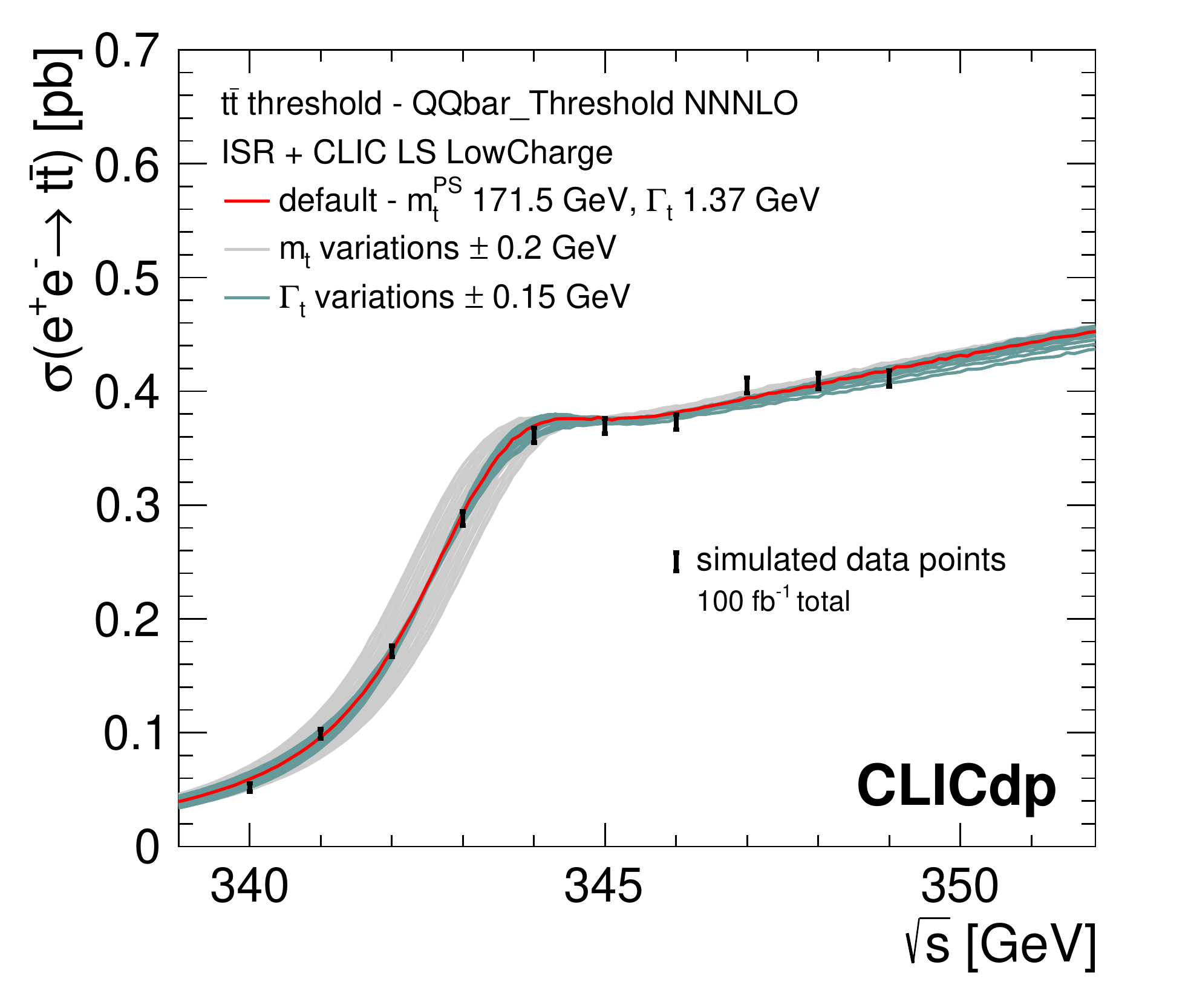}
\end{minipage}
\begin{minipage}{0.51\linewidth}\centering
\includegraphics[width=\textwidth]{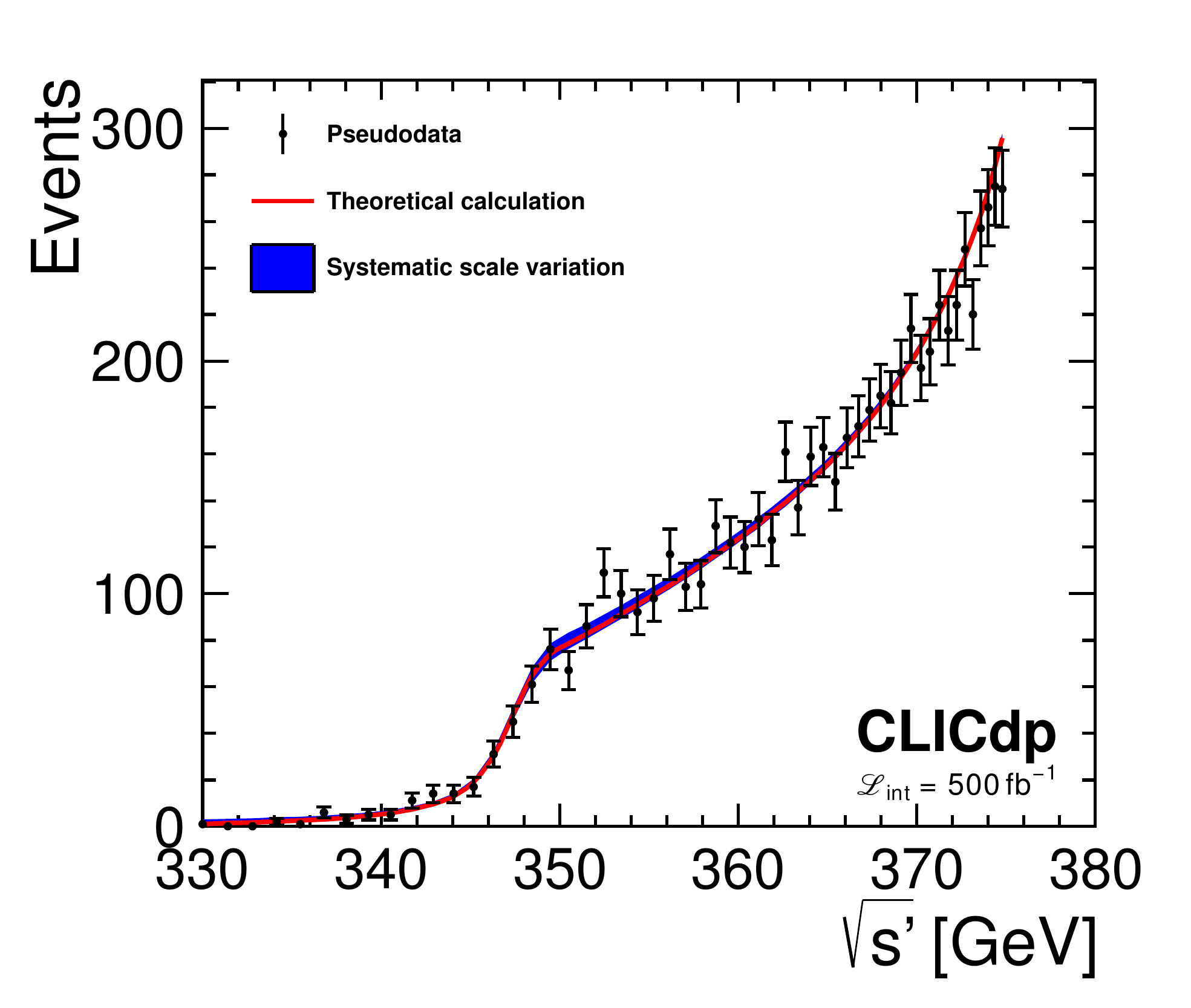}
\end{minipage}
\caption{
  Illustration of measurements allowing a top-quark mass determination at the first stage
  of CLIC: from the dedicated threshold scan (left) and from
  radiative event reconstruction at 380\,GeV (right).
 Figures taken from~\cite{Abramowicz:2018rjq}.
}
\label{fig:mass}
\end{figure}

The most precise determination of the top-quark mass, both in terms of the
statistical and of the systematic uncertainties, is possible with a
dedicated scan of the top pair production threshold.
The shape of the pair-production cross section around the threshold
is very sensitive not only to the top-quark mass, but also to its
width and Yukawa coupling.
Although the dependence is smeared by ISR effects and the beam
luminosity spectrum, 100\,fb$^{-1}$ collected around the threshold with the dedicated
luminosity spectrum is sufficient to reduce the statistical mass
uncertainty (assuming fixed width and Yukawa coupling) to 19\,MeV, see
figure~\ref{fig:mass}\,(left). 
The top-quark mass is extracted in a well-defined theoretical scheme,
significantly reducing the systematic uncertainties.
Combined theoretical uncertainties,
including parametric uncertainty from $\alpha_{s}$
and from QCD scale variations, are estimated to be between 30 MeV and
50 MeV. 
Experimental systematic uncertainties, including
beam energy and the luminosity spectrum, 
selection efficiencies and residual background levels,
are estimated to be at the similar level, between 25 MeV and 50 MeV.
A total systematic uncertainty on the top-quark mass determination from the
threshold scan of around 50 MeV is therefore feasible.

The top-quark mass can also be extracted when running above the
threshold, at 380 GeV, from the cross section of radiative events,
\eettg.
The expected distribution of the reconstructed top-pair invariant mass
is presented in figure~\ref{fig:mass}\,(right).
With the assumed integrated luminosity of 500\,fb$^{-1}$ a
statistical precision of about 100\,MeV can be obtained and 
the total uncertainty is estimated to be about 150\,MeV.

The direct top-quark mass
determination from the reconstructed decay products is accessible at all CLIC energy stages.
Statistical precision of around 40\,MeV is expected at the first CLIC stage
for combined hadronic and semi-leptonic top-quark pair production samples.
To match the statistical accuracy the jet energy scale
should be controlled at the level of 0.025\%.
Jet energy scale effects can be significantly reduced by relating the
top-quark mass to the W boson mass.
The expected statistical uncertainty on the reconstructed mass ratio
corresponds to a top-quark mass uncertainty of about 60 MeV.

\section{Properties of top-quark pair production}

\begin{figure}[p]
\begin{minipage}{0.45\linewidth}\centering
\includegraphics[width=\textwidth]{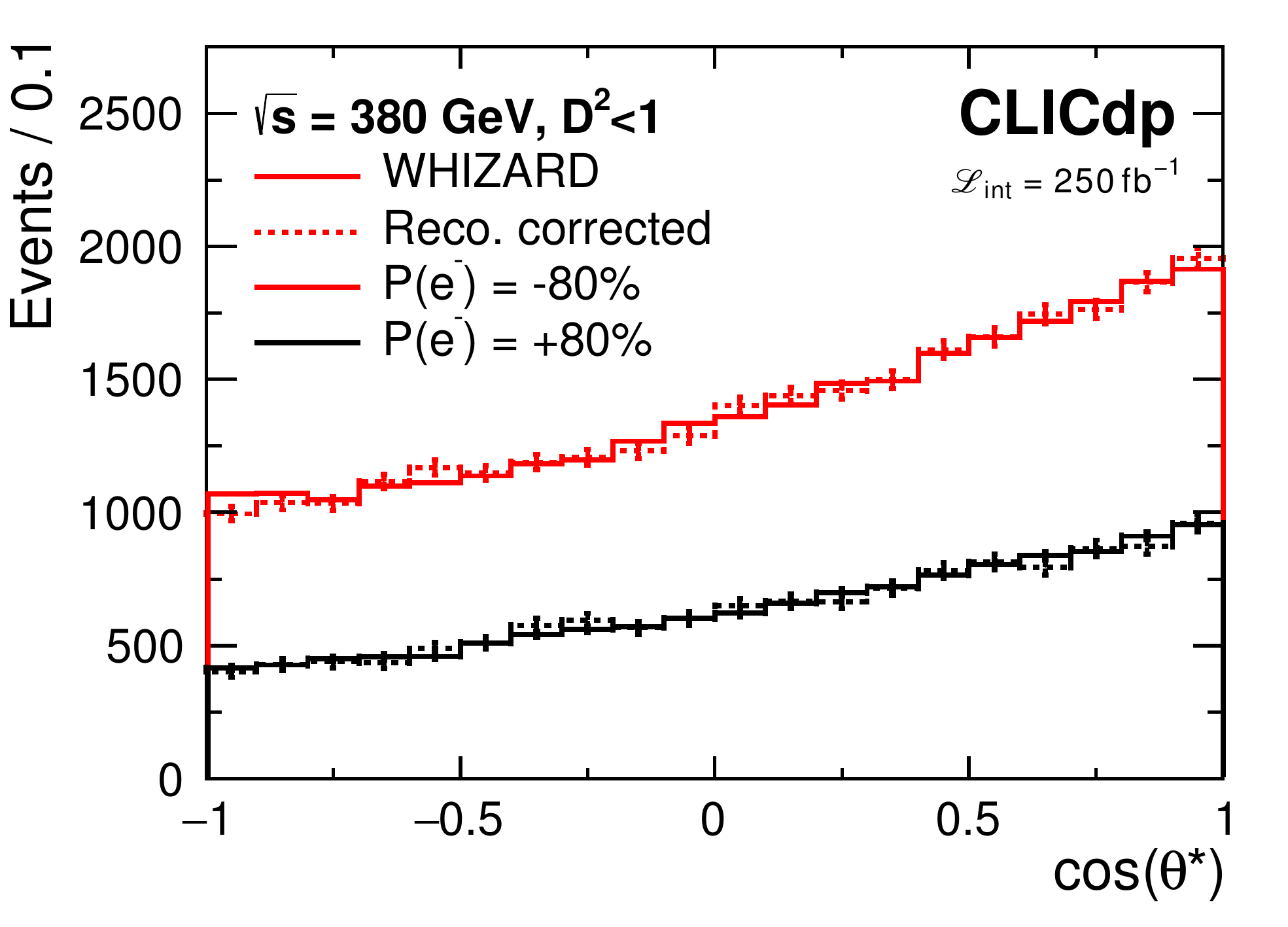}
\end{minipage}
\begin{minipage}{0.55\linewidth}\centering
  \vspace*{-6mm}
\includegraphics[width=\textwidth]{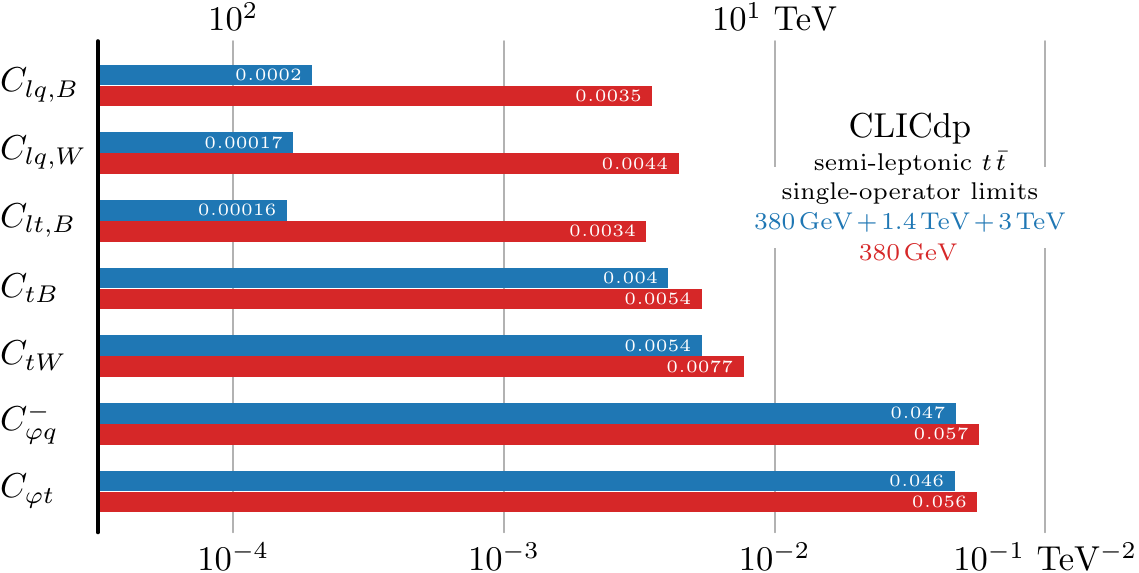}
\end{minipage}
\caption{
  Top-quark polar angle distributions for operation at
  $\sqrt{s}=380$\,GeV after the application of a tight quality cut
  (left) and the expected limits on the  Wilson coefficients from the
  global  EFT analysis of top-quark pair production at the 380\,GeV (right).
  Figures taken from~\cite{Abramowicz:2018rjq}.
  }
\label{fig:coupling}
\end{figure}

Top-quark couplings to the photon and the Z-boson can already be precisely
determined  at the first CLIC stage from measurements of the \ttbar
production cross section and forward-backward asymmetry with
different electron beam polarisations.
The pair production cross section can be measured for both
polarisations with statistical accuracy below 1\% (assuming
250\,fb$^{-1}$  per polarisation) while the forward-backward asymmetry
is extracted from the measured top-quark angular distribution for 
semi-leptonic events with a statistical uncertainty of the order of
4-5\%, see figure~\ref{fig:coupling}\,(left).  
Top-quark measurements can be used to constrain the possible BSM effects
induced by heavy new physics described in terms of Effective Field
Theory (EFT) operators.
Expected limits on the Wilson coefficients for seven EFT operators
contributing to the top-quark pair production are presented in
figure~\ref{fig:coupling}\,(right).
Even at the first CLIC stage mass scales in the 10\,TeV range can
be probed. For four operators, most of the sensitivity is provided by 
the initial 380\,GeV stage.

\section{Flavour-changing neutral current top-quark decays}

FCNC top-quark decays are very strongly suppressed in the SM, with 
expected branching ratios of the order of  $10^{-14}$ to $10^{-12}$.
On the other hand, significant enhancement is expected in many BSM
scenarios, reaching up to $10^{-2}$ for \BR(\tch) and $10^{-5}$ for \BR(\tcg).
Searches for FCNC top decays at CLIC, for channels involving charm
quark, can also be competitive with the expected HL-LHC reach  for these
channels of around $2\cdot 10^{-4}$ \cite{tch@lhc} and $7\cdot 10^{-5}$ \cite{tcg@lhc},
respectively.

Three FCNC decay channels have been studied for the first stage of
CLIC.
All channels profit from the precise final state reconstruction and
high flavour tagging efficiency expected for the CLIC detector.
For \tcg decay we expect a high energy isolated photon and a c-quark jet
in the final state, as well as a b-quark jet and a pair of light jets
from the hadronic decay of the second (``spectator'') top.
The same hadronic final state topology is expected for the \tcx channel, but
with no photon, and large missing energy and momentum from the escaping
massive scalar particle.
Finally, candidate events for \tch decay (followed by Higgs boson decay \hbb)
are selected by looking for a c-quark jet and three b-quark jets, with
invariant mass of two b-quark jets consistent with the H mass.

\begin{figure}[p]
\includegraphics[width=0.49\textwidth]{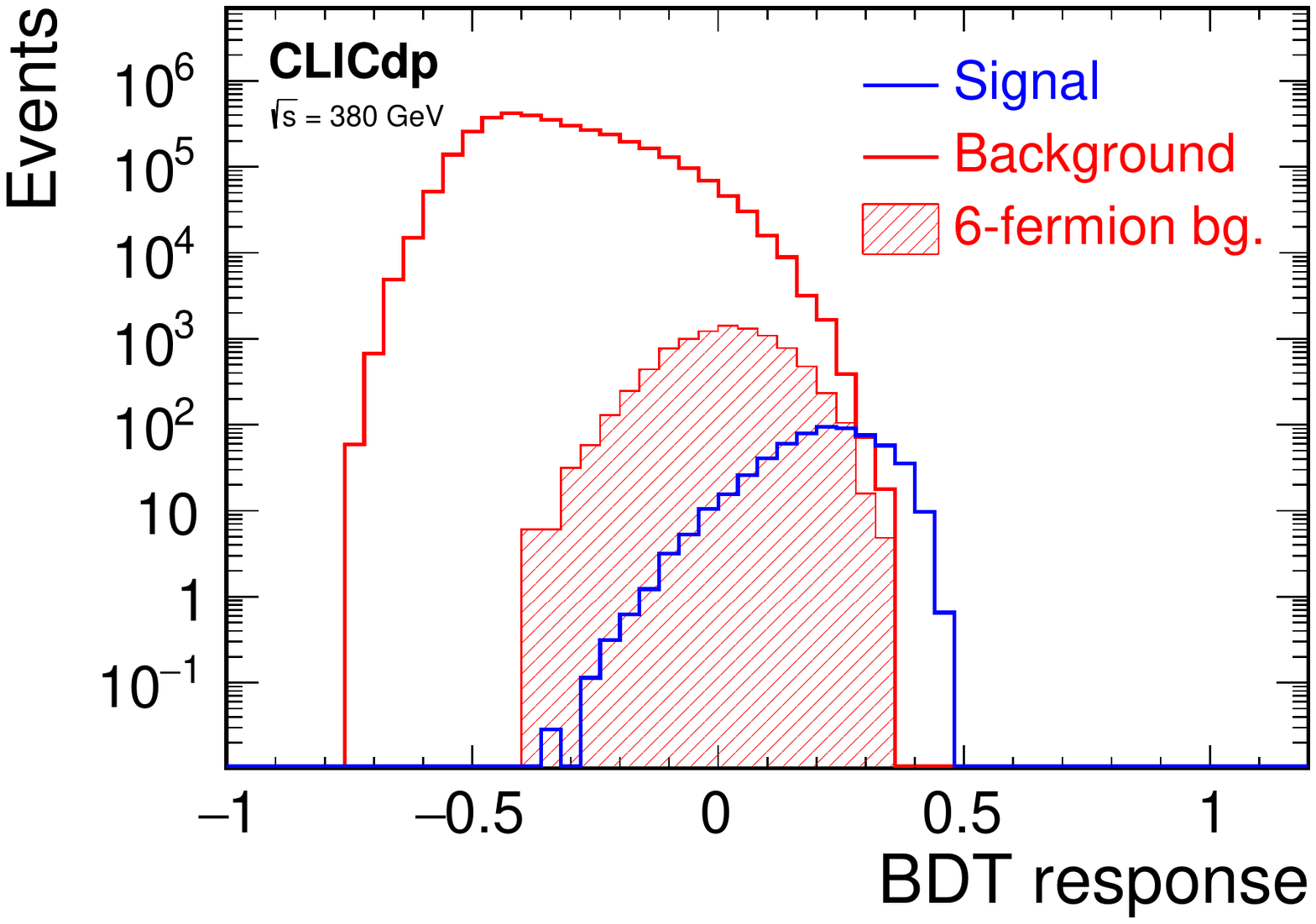}
\includegraphics[width=0.49\textwidth]{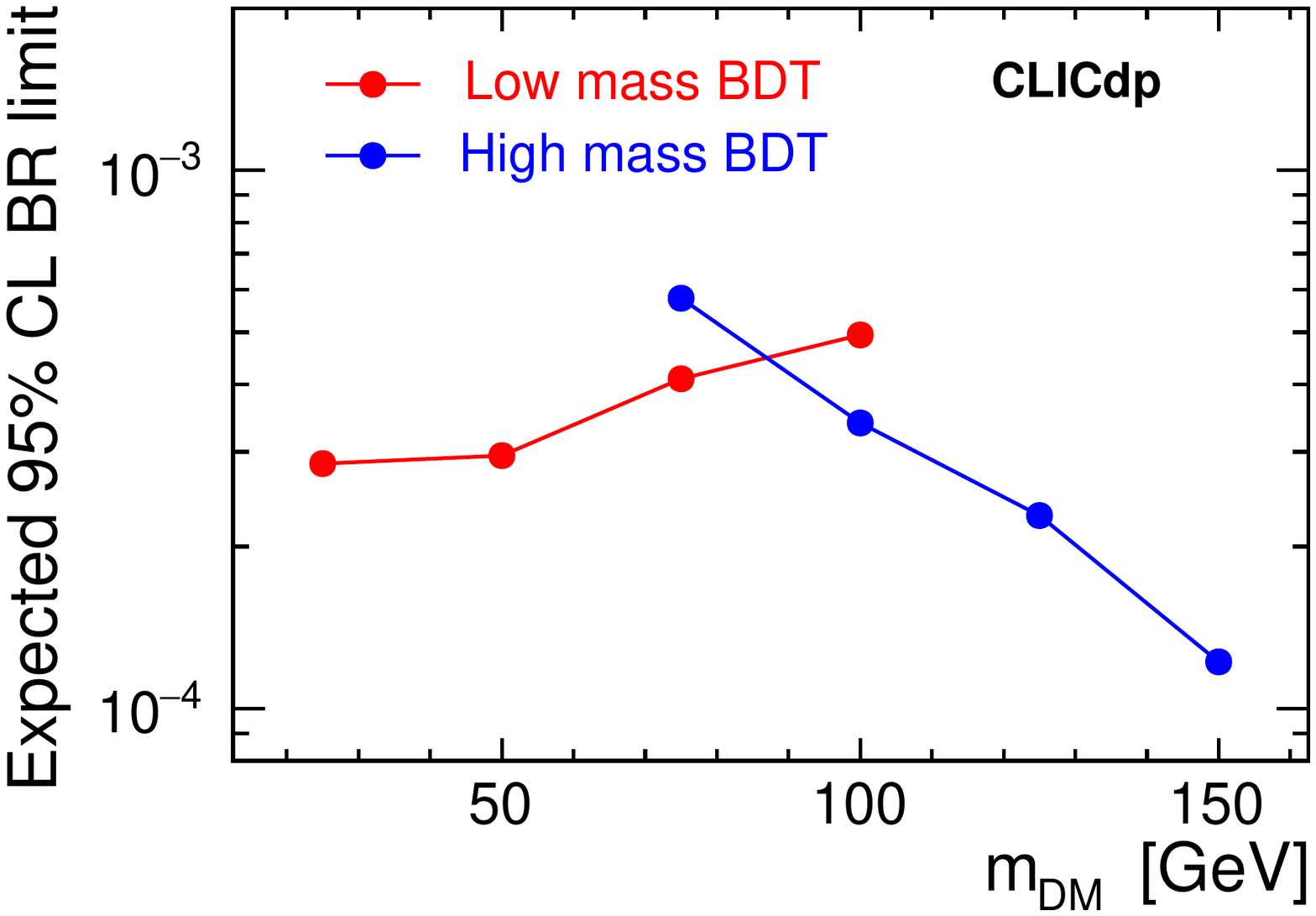}
\caption{
 Distribution of the BDT classifier response for events with 
 FCNC top-quark decay  \tcg and SM background samples (left)
 and the expected 95\% C.L. limits on the top quark FCNC decay \tcx\,
 for 500\,fb$^{-1}$ collected at 380~GeV CLIC (right).
 Figures taken from~\cite{Abramowicz:2018rjq}.
 }
\label{fig:fcnc}
\end{figure}

The same analysis procedure is applied for all channels. Event
pre-selection and classification is based on global event properties,
jet clustering results, flavour tagging as well as lepton and photon
identification and isolation requirements.
A kinematic fit for signal and background hypothesis is then performed
for events matching the required signal final state topology.
Expected limits are then calculated using \CLs approach, based on the
response distribution of the Boosted Decision Tree (BDT) classifier
trained to discriminate between signal and background events.
An example distribution of the BDT classifier response for events with 
FCNC top-quark decay  \tcg and SM background samples is presented in
figure~\ref{fig:fcnc}\,(left).
Expected 95\% C.L. limits for 500 fb$^{-1}$ collected at 380 GeV CLIC
are: $\BR(\tcg)  <  4.7 \cdot 10^{-5}$, 
$\BR(\tch) \times \BR(\hbb)  <   1.2 \cdot 10^{-4}$ and
$\BR(\tcx)  <  1.2 - 4.1 \cdot 10^{-4}$, see figure~\ref{fig:fcnc}\,(right).

\end{document}